\documentclass[prl,reprint,amsmath,amssymb,aps,preprintnumbers,showpacs,twocolumn]{revtex4}
\usepackage[colorlinks=true,linkcolor=black, citecolor=black,urlcolor=black]{hyperref} 
\usepackage{graphicx}  
\usepackage{braket} 
\usepackage{bbm} 
\usepackage{tikz} 

\usetikzlibrary{calc}
\usetikzlibrary{arrows}

\newcommand{\alg}[1]{\mathfrak{#1}}

\newcommand{\su}{\alg{su}}

\newcommand{\ce}{\text{c.e.}}

\def\be{\begin{equation}}
\def\ee{\end{equation}}

\newcommand{\opdtd}{\widetilde{\mathcal{O}}}

\newcommand{\Tr}{\operatorname{Tr}}
\newcommand{\STr}{\operatorname{Str}}
\newcommand{\Ad}{\operatorname{ad}}
\newcommand{\AD}{\operatorname{Ad}}

\newcommand{\dop}{\hat{d}}
\newcommand{\dopt}{\hat{d}^T}
\newcommand{\Pt}{\tilde{P}}

\begin{document}
\preprint{Imperial-TP-RB-2016-06}

\vspace*{0cm}

\title{Integrable deformations of T-dual $\sigma$ models}
\author{Riccardo Borsato}
\email{r.borsato@imperial.ac.uk}
\author{Linus Wulff}
\email{l.wulff@imperial.ac.uk}

\affiliation{The Blackett Laboratory, Imperial College, London SW7 2AZ, United Kingdom}

\begin{abstract}
We present a method to deform (generically non-abelian) T duals of two-dimensional $\sigma$ models, which preserves classical integrability.
The deformed models are identified by a linear operator $\omega$ on the dualised subalgebra, which satisfies the 2-cocycle condition.
We prove that the so-called homogeneous Yang-Baxter deformations are equivalent, via a field redefinition, to our deformed models when $\omega$ is invertible.
We explain the details for deformations of T duals of Principal Chiral Models, and present the corresponding generalisation to the case of supercoset models.
\end{abstract}


\maketitle

\section{Introduction}

Integrable models in two dimensions have played a pivotal role in the understanding of (quantum) field theory, have numerous applications in condensed matter theory, and have recently attracted attention also in the context of the AdS/CFT correspondence~\cite{Maldacena:1997re}, which relates certain string theories on $(d+1)$-dimensional anti de Sitter (AdS) backgrounds to conformal field theories in $d$ dimensions.
The most studied example which exhibits integrable structures is that of the superstring on AdS$_5\times$S$^5$~\cite{Bena:2003wd} and its dual $\mathcal{N}=4$ super Yang-Mills theory in four dimensions~\cite{Minahan:2002ve}, see~\cite{Beisert:2010jr,Bombardelli:2016rwb} for  reviews.
On the string side the two-dimensional worldsheet theory is \emph{classically integrable}, i.e. there is a \emph{Lax pair} whose flatness condition is equivalent to the equations of motion of the $\sigma$ model. The Lax pair depends on an auxiliary spectral parameter $z$, and its expansion around a fixed $z_0$ yields an infinite set of conserved charges, see~\cite{Torrielli:2016ufi} for a review.
Integrability has provided the most stringent tests of AdS/CFT, culminating with the possibility of computing the spectrum of the quantum theory in the large $N$ limit \emph{exactly}~\cite{Bombardelli:2009ns,Gromov:2009bc,Arutyunov:2009ur,Gromov:2013pga}.

Given this tremendous success it is natural to ask whether other theories which are not maximally (super)symmetric are  still integrable.  Integrability could then also be a guiding principle to discover new models which are interesting in their own right.  The $\beta$ deformation~\cite{Lunin:2005jy,Frolov:2005ty,Frolov:2005dj} or certain gravity duals of non-commutative gauge theories~\cite{Hashimoto:1999ut,Maldacena:1999mh} are examples which are integrable but reduce to the maximally symmetric case only when a deformation parameter is sent to zero. These instances actually fall into a larger class that goes under the name of Yang-Baxter (YB) models~\cite{Klimcik:2002zj,Klimcik:2008eq,Delduc:2013qra,Kawaguchi:2014qwa}, sometimes also called $\eta$ deformations after the deformation parameter. A YB model is identified by an $R$ matrix solving the classical Yang-Baxter equation (CYBE), which in general has a rich set of solutions.  Each $R$ generates a background that reduces to the undeformed model (e.g. AdS$_5\times$S$^5$) in the $\eta\to 0$ limit. Here we will not consider the case of ``modified'' CYBE.

In this letter we explore another possibility; we deform the original $\sigma$ model by adding a topological term (a closed B-field) and then apply non-abelian T duality (NATD) \cite{delaOssa:1992vci} with respect to a subgroup $\widetilde G$ of the isometry group $G$. The special case when $\widetilde G$ is abelian gives so-called TsT transformations~\cite{Lunin:2005jy,Frolov:2005ty,Frolov:2005dj}. We refer to the resulting actions as deformed T dual (DTD) models, since sending the deformation parameter $\zeta\to 0$ they reduce to NATD.
DTD models are in one-to-one correspondence with 2-cocycles $\omega$ of the Lie algebra of $\widetilde G$. The cocycle condition~\eqref{eq:cocycle-ggd} guarantees that integrability is preserved, and plays the same role as CYBE for YB models.

The analogy goes even further. When $\omega$ is invertible its inverse $R=\omega^{-1}$ solves CYBE, and  each solution of CYBE corresponds to an invertible 2-cocycle~\cite{STOLIN1999285}. We use this identification to show that the action of YB can be recast in the form of DTD models, where the two deformation parameters are simply related by $\eta=\zeta^{-1}$.
As explained later, this translates into our language a recent conjecture by Hoare and Tseytlin~\cite{Hoare:2016wsk}. We prove it by providing the explicit field redefinition that relates YB to DTD. The field redefinition is \emph{local}, albeit in general non-linear, and it allows us to interpolate between a certain $\sigma$ model ($\zeta\to\infty$) and its NATD ($\zeta\to 0$). In the case when $\omega$ is degenerate, DTD {may be} equivalent to a combination of YB deformation and NATD.

We first construct the DTD of the Principal Chiral Model (PCM), since it provides a simpler set up where all the essential features already appear. Later we generalise it to the case of supercosets, which is more relevant to the study of deformations of superstrings. The supercoset case will be described in more detail elsewhere~\cite{Borsato:2017qsx}.

\section{DTD of PCM}

We start from a PCM parameterised by a group element $g\in G$, with the familiar action $S[g]=-\frac{1}{2}\int  \Tr(g^{-1}\partial_+gg^{-1}\partial_-g)$. Since we want to dualise a  $\widetilde G$ subgroup of the left copy of $G$~\footnote{The construction could be generalised to include the right copy. That DTD should then be related to the bi-YB model of~\cite{Klimcik:2008eq}.} we rewrite~\footnote{We  omit the integration measure $d\sigma^+ d\sigma^-$ where $\sigma^\pm= \tau\pm \sigma$.}
\begin{equation}\label{eq:PCM-action-F+-}
S[f,\tilde A,\nu]=-\tfrac{1}{2}\int  \Tr\left( (\tilde A_+  +J_+)(\tilde A_-  +J_-)+\nu \tilde F_{+-} \right)\,.
\end{equation}
Here $J=dff^{-1}$ is a \emph{right}-invariant Maurer-Cartan form for $f\in G$,  depending on fields that remain \emph{spectators} under NATD.  At the same time $\tilde A\in \tilde{\alg{g}}$  and $\nu\in \tilde{\alg{g}}^*$  identify each of the two T-dual frames. 
If $T_i$ are generators for $\tilde{\alg{g}}$, a basis for the dual algebra $\tilde{\alg{g}}^*$ is given by $T^i$, where  $\Tr(T_iT^j)=\delta_i^j$.
The curvature of $\tilde A$ is $\tilde F_{+-}=\partial_+ \tilde A_--\partial_-\tilde A_+ + [\tilde A_+,\tilde A_-]$. 
The original PCM is recovered upon integrating out $\nu$ since $\tilde F_{+-}=0$ implies that $\tilde A$ is pure gauge, i.e. $\tilde A=\bar g^{-1} d\bar g$ for a  $\bar g\in \widetilde G$, and we get the desired action with $g=\bar g f$. The NATD with respect to $\widetilde G$, on the other hand, is obtained by integrating out $\tilde A$.

We now add a deformation with parameter $\zeta$ given by
\begin{equation}\label{eq:DTD-action-F+-}
S'[f,\tilde A,\nu]=S[f,\tilde A,\nu]+\frac{\zeta}{2}\int  \Tr\left(  \tilde A_+\omega \tilde A_- \right)\,.
\end{equation}
Here $\omega:\,\tilde{\alg{g}}\to \tilde{\alg{g}}^*$ is a linear antisymmetric (i.e. $\Tr(x\omega y)=-\Tr(\omega x y)$) map satisfying the cocycle condition~\footnote{
We use  standard notation $\AD_g M=gMg^{-1}$ and $\Ad_xM=[x,M]$.
Equivalently \eqref{eq:cocycle-ggd} takes the form $\omega(x,[y,z])+\omega(y,[z,x])+\omega(z,[x,y])=0$ for $\omega:\,\tilde{\alg{g}}\otimes \tilde{\alg{g}}\to\mathbb{R}$.}
\be\label{eq:cocycle-ggd}
\omega \Ad_xy=\Pt^T(\Ad_x\omega y-\Ad_y \omega x),\qquad \forall x,y\in \tilde{\alg{g}}\,.
\ee
This property is needed to have local $\widetilde G$ invariance also for $\zeta\neq 0$, which ensures that \# d.o.f.$=\text{dim}(G)$~\footnote{Local invariance is found by including also a shift proportional to $\zeta$ in the transformation for $\nu$. We thank A. Tseytlin for pointing this out.}.
Equations of motion for $\tilde A$ give $\int \Tr ( \delta\tilde A_\mp \mathcal{E}_\pm)=0$ where
\be\label{eq:delta-S-DTD}
\mathcal{E}_\pm\equiv(1\pm \Ad_\nu\pm \zeta \omega)\tilde A_\pm \mp\partial_\pm\nu +J_\pm\,.
\ee
This implies $\Pt^T\mathcal{E}_\pm=0$, where $\Pt$ projects onto $\tilde{\alg{g}}$, $\Pt^T$ onto $\tilde{\alg{g}}^*$. 
We  solve these equations by defining the linear operator $\opdtd =\Pt^T( 1- \Ad_\nu- \zeta \omega)\Pt$ which is a map $\tilde{\alg{g}}\to \tilde{\alg{g}}^*$
\begin{equation}\label{eq:sol-A+-}
\begin{aligned}
 \tilde A_- = \opdtd^{-1} \left(-\partial_-\nu -J_-\right),
\ \ 
 \tilde A_+ = \opdtd^{-T} \left(\partial_+\nu -J_+\right)
\end{aligned}
\end{equation}
and $\opdtd^{-T}$ is the inverse of its transpose. Note that $\opdtd^{-1}\opdtd=\Pt$ as the LHS is defined only on $\tilde{\alg{g}}$.
Evaluating $S'$ on the solution we get the DTD action
\begin{equation}\label{eq:DTD-action}
\begin{aligned}
S'[f,\nu]=-\tfrac{1}{2}&\int \Tr\Big( J_+J_-\\
&+(\partial_+\nu -J_+)\opdtd^{-1}(\partial_-\nu +J_-)\Big)\,.
\end{aligned}
\end{equation}
A second interpretation of DTD comes from integrating out $\nu$ rather than $\tilde A$ from~\eqref{eq:DTD-action-F+-}, which gives again $\tilde A=\bar{g}^{-1}d\bar g$. The resulting action is a \emph{topological} deformation of the PCM, since the cocycle condition implies that $B=\zeta\omega(\bar{g}^{-1}d\bar g,\bar{g}^{-1}d\bar g)$ is closed. At the classical level adding this term has no effect, and in fact this picture of a deformation which is trivial in the dual frame is reminiscent of YB models: in some cases they correspond to TsT transformations~\cite{Orlando:2016qqu,Osten:2016dvf,Borsato:2016ose,Hoare:2016wsk}, which are just field redefinitions in a T-dual frame.
Since DTD is a NATD of a topological deformation of PCM, it is classically integrable, where  NATD can be applied thanks to closure of $B$.
In fact, the equation $\tilde A=\bar{g}^{-1}d\bar g$ with $\tilde A$ given in~\eqref{eq:sol-A+-} allows us to relate the variables of the deformed model to those of the original PCM. In the special case of abelian subalgebra $\tilde{\alg{g}}$ the relation simplifies and the deformed model becomes equivalent to the PCM with twisted boundary conditions, consistent with the TsT interpretation~\cite{Frolov:2005ty}.


A third interpretation of DTD comes from the possibility of applying NATD to a \emph{centrally extended} subalgebra. This idea first appeared in~\cite{Hoare:2016wsk} and was the original motivation for considering the deformation (\ref{eq:DTD-action-F+-}). One can indeed replace $\tilde A$ in~\eqref{eq:PCM-action-F+-} with $\tilde A'\in\tilde{\alg{g}}_{\ce}=\tilde{\alg{g}}\oplus \alg{c}$ and $\alg{c}$ central;  similarly $\nu'\in \tilde{\alg{g}}_{\ce}^*$. We decompose $\tilde A'=\tilde A+\tilde A^{\alg{c}},\ \nu'=\nu+\nu^{\alg{c}}$ with obvious notation, and extend the definition of the trace $\Tr(\alg{c}^2)=1,\ \Tr(\alg{c}\alg{g})=0$.
Equations for $\tilde A^{\alg{c}}$ imply that $\nu^{\alg{c}}$ is constant,  $\nu^{\alg{c}}=\zeta \alg{c}$. At this point $\Tr(\nu' \tilde F'_{+-})=\Tr(\nu \tilde F_{+-})+\zeta \mathbf{f}_{ab}\tilde A^a_+\tilde A^b_-$, where $\mathbf{f}_{ab}$ are the  structure constants introduced by the central extension $[T_a,T_b]=f_{ab}^c T_c+\mathbf{f}_{ab}\alg{c}$. Introducing a map $\omega$ whose components are $\omega_{ab}=-\mathbf{f}_{ab}$ we just notice that it is antisymmetric and satisfies  the cocycle condition, a consequence of the Jacobi identity in $\tilde{\alg{g}}_{\ce}$ projected on $\alg{c}$. 

For some $\omega$'s DTD reduces to just NATD, i.e. the deformation parameter can be removed by a field redefinition.
This happens when $\omega$ is a coboundary, i.e. $\omega(x,y)=f([x,y])$ for some function $f$.
Therefore, non-trivial deformations are in one-to-one correspondence with 2-cocycles modulo coboundaries, i.e. with elements of the second cohomology group $H^2(\tilde{\alg{g}})$. The same holds also for  non-trivial central extensions. In particular, there are none for semisimple $\tilde{\alg{g}}$.
Trivial deformations are equivalently described as adding an \emph{exact} B-field to PCM.

\section{An example}

Before continuing our general discussion, let us  provide an explicit example: a PCM on $U(2)$. We use generators  $T_j=i\sigma_j\in\su(2)$ and $T_4=i \mathbf{1}$, with duals   $T^j=-\tfrac{i}{2}\sigma_j$ and $T^4=-\tfrac{i}{2} \mathbf{1}$. We  parameterise the group element by $g=\exp(i \theta \mathbf{1})\exp(i\phi_+ \sigma_1)\check g(\xi)\exp(i\phi_- \sigma_2)$, where $\phi_\pm=(\phi_1\pm\phi_2)/2$ and $\check g(\xi)=\text{diag}(i^{-1/2}e^{i\xi},i^{1/2}e^{-i\xi})$. The PCM action yields the metric of S$^3\times$S$^1$
\be
ds^2=d\xi ^2+\sin ^2\xi \  d\phi_1^2+\cos ^2\xi\  d\phi_2^2 +d\theta^2\,.
\ee
Suppose we want to dualise the coordinates $\phi_+$ in S$^3$ and $\theta$ in S$^1$, corresponding to the  abelian subalgebra $\tilde{\alg{g}}=\text{span}\{T_1,T_4\}$. We  take $f=\check g(\xi)\exp(i\phi_- \sigma_2)$ and $\nu=2(\tilde\phi_+ T^1+\tilde \theta T^4)$, where $\tilde\phi_+,\tilde \theta$ are dual coordinates.
We  deform the dual theory by taking $\omega = 2 T^1\wedge T^4$, namely $\omega T_1=-2T^4,\ \omega T_4=2T^1$. From~\eqref{eq:DTD-action} we find the action of DTD $S'=\int \partial_+X^i(G_{ij}-B_{ij})\partial_-X^j$, with the  metric and B-field
\be
\begin{aligned}
&ds^2=d\xi^2+(1+\zeta^2)^{-1}\Big( d\tilde{\phi}_+^2+ \left(\zeta ^2+\sin ^22 \xi \right)\ d\phi_-^2\\
&\qquad\quad\quad+d\tilde{\theta}^2+2 \zeta   \cos 2 \xi \ d\tilde{\theta}d\phi_-\Big)\,,\\
&B=(1+\zeta^2)^{-1}\Big(\cos 2\xi \ d\phi_--\zeta \ d\theta\Big)\wedge d\tilde\phi_+\,.
\end{aligned}
\ee
The $\zeta \to 0$ limit yields the T-dual model of S$^3\times$S$^1$ with respect to  $\tilde{\alg{g}}$.
To relate this simple example to a YB model it is enough to take $\nu=\eta^{-1} R(\vartheta T^4+\varphi_+T^1)$ with $R=\tfrac{1}{2}(T_4\wedge T_1)$. However, when  $\tilde{\alg{g}}$ is non-abelian, the field redefinition is more complicated, see~\eqref{eq:red-nu-X}.

\section{Integrability}

Above we argued that DTD models must be integrable, however it is instructive to show this explicitly to see how the cocycle condition enters and  write a Lax connection. We will show that the equations of motion  formally resemble those of the PCM, for which a Lax pair is known. 
Suppose we consider a PCM with group element $g=\bar g f$, with $\bar g\in \widetilde G,\ f\in G$.
We prefer to rewrite its on-shell equations in terms of the left and right currents $\tilde A=\bar g^{-1} d\bar g$ and $J=df f^{-1}$. To start, the flatness condition for $A=g^{-1}dg$ is equivalent to $\mathcal{F}^J=0,\ \mathcal{F}^{\tilde A}=0$
\be
\begin{aligned}
\mathcal{F}^J&\equiv \partial_+J_--\partial_-J_+-[J_+,J_-],\\
\mathcal{F}^{\tilde A}&\equiv\partial_+\tilde A_--\partial_-\tilde A_++[\tilde A_+,\tilde A_-].
\end{aligned}
\ee
Moreover, the equations of motion for the PCM, i.e. conservation of $A$, become $\mathcal{C}=0$
\be
\mathcal{C}\equiv\partial_+(J_-+\tilde A_-)+\partial_-(J_++\tilde A_+)+[\tilde A_+,J_-]+[\tilde A_-,J_+].
\ee
Let us now rederive the above equations for DTD models, where now importantly $\tilde A$ is identified as in~\eqref{eq:sol-A+-}. To start, the flatness condition $\mathcal{F}^J=0$ still follows from the definition of $J$. 
Flatness for $\tilde A$, instead, now arises as the equations of motion for $\nu$ which are $\delta_\nu S'[f,\nu]=-\frac{1}{2}\int  \Tr(\delta \nu\ \mathcal{F}^{\tilde A})=0$.
It is nice that the known  mechanism familiar from T duality of trading flatness for an equation of motion  still holds for DTD.

Equations of motion for $f$ are $\delta_f S'[f,\nu]=+\frac{1}{2}\int \Tr( \delta ff^{-1}\ \mathcal{C}) =0$, essentially as in the previous example of PCM. However, in that case it is only thanks to the equations of motion for $\bar g$ (i.e. $\int\Tr( \bar g^{-1} \delta \bar g\ \mathcal{C}) =0$) that one can claim  $\mathcal{C} =0$. In analogy to PCM, it is then clear that our task is to show that $\Pt^T\mathcal{C}=0$ also for DTD.
We generalise the argument of~\cite{Sfetsos:2013wia} for NATD of PCM, and consider the equations $\mathcal{E}_\pm=M^\perp_\pm$, for some $M^\perp_\pm$ for which $\Pt^TM^\perp_\pm=0$. They imply $\Pt^T \mathcal{E}_\pm=0$, i.e. they are equivalent to the solutions for $\tilde A$ as in~\eqref{eq:sol-A+-}. They obviously imply also the equation $(\partial_++\Ad_{\tilde A_+})(\mathcal{E}_--M^\perp_-)+(\partial_-+\Ad_{\tilde A_-})(\mathcal{E}_+-M^\perp_+)=0$, which  reads as
$$
\begin{aligned}
\mathcal{C}&=[\partial_-+\Ad_{\tilde A_-},\partial_++\Ad_{\tilde A_+}]\nu\\
&-(\partial_-+\Ad_{\tilde A_-})M^\perp_+-(\partial_++\Ad_{\tilde A_+})M^\perp_-\\
&+\zeta(\omega (\partial_+\tilde A_--\partial_-\tilde A_+)+\Ad_{\tilde A_+}\omega\tilde A_--\Ad_{\tilde A_-}\omega\tilde A_+)\,.
\end{aligned}
$$
The first line on the right hand side is rewritten as $[\nu,\tilde F_{+-}]$, and hence vanishes thanks to flatness of $\tilde A$. The second line vanishes upon projecting with $\Pt^T$~\footnote{If $\Pt^T(M^\perp_\pm)=0$ then also $\Pt^T(\Ad_xM^\perp_\pm)=0$ with $x\in\tilde{\alg{g}}$.}.
Finally, the last line vanishes thanks to the cocycle condition: using~\eqref{eq:cocycle-ggd} it is rewritten as $-\zeta\omega(\tilde F_{+-})$, which is again zero. Since also $\Pt^T\mathcal{C}=0$ holds, we conclude that the whole set of on-shell equations for the DTD is formally equivalent to those of a PCM, provided the proper $\tilde A$ is used. 
We can furthermore write the Lax pair as
\be\label{eq:Lax-PCM-split}
L_\pm=\tfrac{1}{2}(1+z^{\mp 2})\AD_{f}^{-1}(\tilde A_\pm+J_\pm)\,,
\ee
with $z$ a  spectral parameter. In fact, the flatness condition
$\partial_+L_--\partial_-L_++[L_+,L_-]=0$ is equivalent to the on-shell equations  just derived.

\section{Relation to Yang-Baxter}

We now prove that YB deformations for PCM on the group $G$ are equivalent to DTD. This was checked for many particular examples  in~\cite{Hoare:2016wsk}. YB models are identified by an $R$ matrix solving the CYBE on the Lie algebra $\alg{g}$.
If $g\in G$
\be\label{eq:eta-PCM}
S_{\text{YB}}[g]=-\tfrac{1}{2}\int \Tr\left(g^{-1}\partial_+g\frac{1}{1-\eta R_g}  g^{-1}\partial_-g\right)\,.
\ee
$R$ is invertible on a certain subalgebra and its inverse is a 2-cocycle~\cite{STOLIN1999285}. As anticipated, we  identify $R=\omega^{-1}$, where $\omega$ is the operator defining the DTD model. Then $R:\tilde{\alg{g}}^*\to \tilde{\alg{g}}$. The two deformation parameters will be related by $\eta=\zeta^{-1}$.

We first split the group element parameterising the YB model as $g=\tilde g f$, where $\tilde g\in \widetilde G$ and $f\in G$.
We identify $f$ with the homonym appearing on the DTD side. Our proof of equivalence of the two actions will then consist in giving the field redefinition relating $\tilde g$ and $\nu$.
Since $R$ is invertible, we can always take $\tilde g=\exp(RX)$ for some $X\in \tilde{\alg{g}}^*$. One can check that taking $X=\eta\nu+\frac{\eta^2}{2}\Pt^T[R\nu,\nu]+\mathcal{O}(\eta^3)$ the two actions are equivalent up to terms which are at least cubic in $\eta$. The generalisation to all orders can be obtained by requiring that the $dfdf$ terms in the two actions match. This leads to the condition $(1-\eta  R_{\tilde g})^{-1}=1-\opdtd^{-1}$ whose solution can be shown to be
\begin{equation}\label{eq:red-nu-X}
\nu=\frac{1}{\eta}\tilde P^T\frac{1-e^{-\Ad_{RX}}}{\Ad_{RX}}X=\frac{1}{\eta}\tilde P^T\frac{1-\AD_{\tilde g}^{-1}}{\log\AD_{\tilde g}}\omega\log\tilde g\,.
\end{equation}
It follows  that $d\nu=(\Pt^T-\opdtd)\tilde g^{-1}d\tilde g$ or, equivalently,
\be
\mathbf{ A}_\pm=\AD_{f}^{-1}(J_\pm+\tilde A_\pm)\,,
\ee
where we defined $\mathbf{A}_\pm=(1\pm \eta R_g)^{-1}(g^{-1}\partial_\pm g)$ on the YB side. Using these relations it is not hard to check that the two actions are the same \emph{up to} the topological term  $\zeta\omega(\tilde g^{-1}d\tilde g,\tilde g^{-1}d\tilde g)$, which has no effect in the classical theory as remarked earlier.

We have proven the equivalence of DTD and YB when $\omega$ is non-degenerate. In the case of degenerate $\omega$ it is {often \footnote{In earlier versions it was incorrectly claimed that this is always possible.}} possible to choose it in such a way that it is non-degenerate on a subalgebra $\hat{\alg{g}}\subset\tilde{\alg{g}}$ 
 and acts trivially on its complement $\check{\alg{g}}$ in $\tilde{ \alg{g}}$, also an algebra thanks to~\eqref{eq:cocycle-ggd}. {It would be natural to interpret this} as NATD on $\check{ \alg{g}}$ of the YB model corresponding to restricting $\omega$ to $\hat{\alg{g}}$.

\section{DTD of Supercosets}

The construction of DTD for supercosets follows the steps explained in the simpler case of PCM. Here we  only present the main results, whose derivation will be collected in~\cite{Borsato:2017qsx}.

We still denote by $G$ the group of superisometries, e.g. $PSU(2,2|4)$ for superstrings on AdS$_5\times$S$^5$, see~\cite{Arutyunov:2009ga} for a review. Its Lie superalgebra $\alg{g}$ admits a $\mathbb{Z}_4$ decomposition, and we denote by $P^{(j)}$ the projectors onto the four subspaces. They typically appear in the combination $\dop=P^{(1)}+2P^{(2)}-P^{(3)}$ or its transpose $\dopt$. The absence of $P^{(0)}$ in $\dop$ is necessary for the local $\alg{g}^{(0)}$ invariance of the action, i.e. local Lorentz transformations. The action for DTD of supercosets is~\footnote{We have fixed conformal gauge, $\gamma^{+-}=\gamma^{-+}=\epsilon^{-+}=-\epsilon^{+-}=2$.}
\begin{equation}\label{eq:DTD-supcos-action}
\begin{aligned}
S'[f,\nu]=-&\frac{T}{2}\int  \STr\Big( J_+\dop_fJ_-\\
&+(\partial_+\nu -\dopt_fJ_+)\opdtd_-^{-1}(\partial_-\nu +\dop_fJ_-)\Big),
\end{aligned}
\end{equation}
where $\dop_f\equiv\AD_{f}\dop\AD_{f}^{-1}$. We keep the same definitions for $J,\nu$, which however  now  take values in superalgebras.
Moreover now $\opdtd=\Pt^T(\dop_f-\Ad_\nu -\zeta\omega )\Pt$.

The  model is integrable since we can write down a Lax pair. This is more conveniently expressed in terms of $A=\AD_{f}^{-1} (\tilde A  +J)$, where
\be
\begin{aligned}
\tilde A_+&=\opdtd^{-T}(+\partial_+ \nu -\dopt_fJ_+),\\
\tilde A_-&=\opdtd^{-1}(-\partial_- \nu -\dop_fJ_-).
\end{aligned}
\ee
Then  flatness condition $\partial_+\mathcal{L}_--\partial_-\mathcal{L}_++[\mathcal{L}_+,\mathcal{L}_-]=0$ for
\be
\mathcal{L}_\pm=A^{(0)}_\pm+zA^{(1)}_\pm+z^{\mp 2}A^{(2)}_\pm+z^{-1}A^{(3)}_\pm,
\ee
is equivalent to the on-shell equations of the DTD model.

DTD of supercosets possess kappa symmetry, and therefore correspond to solutions of the generalised supergravity equations of~\cite{Arutyunov:2015mqj,Wulff:2016tju}. Kappa symmetry transformations are  $\delta ff^{-1}=\dopt_f(\delta \nu)=\rho_{1,-}+\rho_{3,+}$, where
\be
\rho_{j,\pm}=\{i\AD_f\kappa^{(j)},J^{(2)}_\pm+\tilde A_\pm^{(2)}\},
\ee
and $\kappa^{(j)},\ j=1,3$ are two local parameters of grading $j$. The action~\eqref{eq:DTD-supcos-action} is invariant under these transformations upon using the Virasoro constraints. If we were not fixing conformal gauge, the variation of the action  would be compensated by the variation of the worldsheet metric.
From these kappa symmetry transformations it is possible to extract the background fields of DTD~\cite{Borsato:2017qsx}.

The equivalence to YB for invertible $\omega$'s holds also in the case of DTD of supercosets. Remarkably, the field redefinition  is still given by~\eqref{eq:red-nu-X} as for PCM. We have further verified that  kappa symmetry transformations of YB models~\cite{Delduc:2013qra} take the above form under this field redefinition, when we fix the $\widetilde G$  gauge to get $\delta ff^{-1}=\dopt_f(\delta \nu)$.

\section{Conclusions}
We provided a unified picture of (non-abelian) T duality and homogeneous YB deformations as DTD of $\sigma$ models. As  pointed out in~\cite{Hoare:2016wsk}, an advantage of this formulation is that it  can be realised at the path integral level, giving a  better handle on the quantum theory. In fact,  it also explains why the condition for one-loop Weyl-invariance, i.e. unimodularity of $\tilde{\alg{g}}$, is the same for both YB model and NATD~\cite{Elitzur:1994ri,Alvarez:1994np,Borsato:2016ose}. 

Despite the close relation, it is still worth to view DTD as a distinct class of deformations. In fact, the field redefinition that relates it to YB is singular in the two undeformed limits; YB becomes degenerate when taking the undeformed (i.e. $\zeta\to 0$) limit of DTD, and viceversa. Therefore, the interpretation as deformation applies to just one of the two models in the T-dual pair.
It would be interesting to understand if there is any connection to the $\lambda$-model of~\cite{Sfetsos:2013wia,Hollowood:2014rla,Hollowood:2014qma}, which is also a deformation of NATD and is related to the inhomogeneous YB deformation~\cite{Klimcik:2002zj,Klimcik:2008eq,Delduc:2013qra}.

Although our  motivation was integrability, such deformations can be applied also to non-integrable models, which provides  an interesting and potentially useful way to generate new supergravity solutions.

\section{Acknowledgements}
We thank Ben Hoare, Stijn van Tongeren and Arkady Tseytlin for interesting discussions and comments on the manuscript. RB  thanks also Bogdan Stefa\'nski for related discussions. RB thanks also Wim Hennink and his group in Utrecht for the kind hospitality during part of this project.
This work was supported by the ERC Advanced grant No.~290456. The work of LW was also supported by the STFC Consolidated grant  ST/L00044X/1.

\bibliographystyle{h-physrev}
\bibliography{biblio}

\end{document}